\documentclass{PoS}

\usepackage{amsmath}		
\usepackage{bm}					

\newcommand{\fig}[1]{Fig.~\ref{#1}}
\newcommand{\eq}[1]{Eq.(\ref{#1})}
\newcommand{\sect}[1]{Section~\ref{#1}}

\title{Automation of 2-loop Amplitude Calculations}
\ShortTitle{Automation of 2-loop Amplitude Calculations}

\author{
\speaker{S.~P.~Jones} \\
Max Planck Institute for Physics, F\"ohringer Ring 6, 80805 Munich, Germany\\
E-mail: \email{sjones@mpp.mpg.de}
}

\abstract{
Some of the tools and techniques that have recently been used to compute Higgs boson pair production at NLO in QCD are discussed.  
The calculation relies on the use of integral reduction, to reduce the number of integrals which must be computed, and expressing the amplitude in terms of a quasi-finite basis, which simplifies their numeric evaluation.
Emphasis is placed on sector decomposition and Quasi-Monte Carlo (QMC) integration which are used to numerically compute the master integrals. 
}

\FullConference{Loops and Legs in Quantum Field Theory\\
		 24-29 April 2016\\
		 Leipzig, Germany}

\begin{document}

\section{Introduction}

Progress in precision calculations is driven primarily by the development of new methods of computation.
However, especially beyond NLO, these calculations are usually performed with some degree of automation.
Here we discuss progress in developing an automated framework for precision calculations which can compute amplitudes numerically. This is particularly useful when the integrals appearing in a process of interest can not be easily obtained analytically.

One important component of this framework is the use of sector decomposition and QMC integration.
Previously it has been shown that QMC integration can perform well when applied to sector decomposed integrals~\cite{Li:2015foa}. Following this work, we have implemented such an integrator and applied it to sector decomposed (mostly) quasi-finite integrals to compute, for the first time, the Higgs boson pair production differential cross-section at NLO~\cite{Borowka:2016ehy}.  

After providing an overview of our set-up we review sector decomposition in \sect{sec:sectordecomposition}, which is used to prepare the integrals for numerical integration. QMC integration methods and some details of our implementation are described in \sect{sec:r1sl}. We also describe a standard procedure which was used to reduce the number of phase-space points required. Although the techniques are not specific to a given process, where it is helpful to do so, we will refer to their application to the $gg \rightarrow HH$ calculation. Results and further discussion of other important techniques used in this calculation have been presented in Ref.~\cite{mkproceedings}.

\section{Overview of Method \& Tool Chain}

Our calculation of the $gg \rightarrow HH$ differential cross-section at NLO primarily uses standard loop techniques. We use conventional dimensional regularisation (CDR) with $D=4-2\epsilon$. Before contracting the amplitude with the gluon polarisation vectors, $\mathcal{M} = \epsilon^1_\mu \epsilon^2_\nu \mathcal{M^{\mu \nu}}$, it can be decomposed into a sum of form factors $F_1, F_2$, which contain the loop integrals, multiplied by known basis tensors $T_1^{\mu \nu}, T_2^{\mu \nu}$ as was done for the original LO calculation~\cite{Glover:1987nx},
\begin{equation}
\mathcal{M}^{\mu \nu} = F_1( \hat{s}, \hat{t}, m_h^2, m_t^2, D) T_1^{\mu \nu} + F_2(\hat{s}, \hat{t}, m_h^2, m_t^2, D) T_2^{\mu \nu}.
\end{equation}
$D$-dimensional projectors $P_1^{\mu \nu}, P_2^{\mu \nu}$ are then constructed such that
\begin{align}
\begin{split}
P_1^{\mu \nu} \mathcal{M}_{\mu \nu} &= F_1( \hat{s}, \hat{t}, m_h^2, m_t^2, D), \\
P_2^{\mu \nu} \mathcal{M}_{\mu \nu} & = F_2( \hat{s}, \hat{t}, m_h^2, m_t^2, D). \label{eq:proj}
\end{split}
\end{align}
These projectors are currently not determined automatically by our set-up and must be input by hand.

Next, we generate Feynman diagrams using \texttt{QGRAF}~\cite{Nogueira:1991ex}, insert the Feynman rules, contract the amplitude with the projectors as in \eq{eq:proj} and renormalize. This procedure is driven by an extended version of \texttt{GOSAM}~\cite{Cullen:2011ac,Cullen:2014yla} which uses \texttt{FORM}~\cite{Vermaseren:2000nd,Kuipers:2012rf} for algebraic manipulations. The amplitude was also generated directly in \texttt{REDUZE}~\cite{vonManteuffel:2012np} and cross-checked prior to integral reduction. The challenging task is to manipulate and calculate the resulting amplitude, which contains many 2-loop integrals depending on up to 4 scales.

To reduce the number of integrals we use integral reduction~\cite{Tkachov:1981wb,Chetyrkin:1981qh,Gehrmann:1999as,Laporta:2001dd} as implemented in \texttt{REDUZE}. Integral reduction can be an extremely computationally expensive task (both in terms of operations and memory usage) and the choice of integral families can greatly affect the resources required to compute a reduction. It is not currently easy to generate `good' integral families which make obtaining an integral reduction simpler or even feasible\footnote{Choosing integral families with permutation symmetries can help. Algorithms to find families with a maximal number of permutation symmetries for massless integrals are available in \texttt{REDUZE}.}.
In fact, for this project already, a reduction could not always be obtained. Specifically, we were so far unable to obtain the reduction for non-planar 4-point integrals, which start at 6 propagators, with \texttt{FIRE}~\cite{Smirnov:2014hma}, \texttt{Litered}~\cite{Lee:2013mka} or \texttt{Reduze}. In this case the scalar integrals with inverse propagators were rewritten in terms of tensor integrals, 
sector decomposed and then integrated numerically using the techniques described below. Currently, integral families must be input into our framework by hand.

We faced various 2-loop integrals which are not known analytically. For their evaluation we applied sector decomposition and numerically integrated the resulting functions using a QMC algorithm. The application of sector decomposition to all integrals appearing and the generation of a single code to evaluate the full renormalized 2-loop amplitude utilising (optionally) multi-threading and/or GPUs is automatic in our set-up.

Finally, parton level events are generated, computed, convoluted with parton distribution functions and used to obtain a (differential) cross-section.

The techniques presented in this section and their implementation in our framework are not specific to the $gg \rightarrow HH$ calculation and can be applied to other processes. However, since Higgs boson pair production is a loop induced process, in this case the computation of the real radiation and IR subtraction could be done using the methods developed for NLO calculations~\cite{Catani:1996vz}. The computation of the renormalized amplitude, expanded in $\epsilon$, was done separately and then combined with the other parts of the calculation using Catani-Seymour dipole subtraction. For the real radiation we used \texttt{GOSAM} to generate the required matrix elements. 

\section{Sector Decomposition} \label{sec:sectordecomposition}

To compute the integrals we utilise sector decomposition~\cite{Hepp:1966eg,Roth:1996pd,Binoth:2000ps,Heinrich:2008si}. Here we briefly outline the procedure and recall the structure of the resulting functions which are later numerically integrated using the method described in \sect{sec:r1sl}.

A generic $L$-loop integral with $N$ propagators, $P_j$, raised to arbitrary powers, $\nu_j$, can be written as
\begin{align}
G 
& = \int_{- \infty}^{\infty} \left( \prod_{l=1}^L \frac{\mathrm{d}^D k_l }{i\pi^\frac{D}{2}} \right) \frac{1}{\prod_{j=1}^N P_j^{\nu_j} } \nonumber \\
& = (-1)^{N_\nu} \frac{\Gamma(N_\nu-LD/2)}{\prod_{j=1}^N \Gamma(\nu_j)} \int_0^\infty \left(\prod_{j=1}^N \mathrm{d}x_j \ x_j^{\nu_j-1} \right) \delta(1-\sum_{i=1}^N x_i)
\frac{\mathcal{U}^{N_\nu - (L+1)D/2}(x_1,\ldots,x_N)}{\mathcal{F}^{N_\nu-LD/2}(x_1,\ldots,x_N; s_1,\ldots,s_m)}. \label{eq:feynmanintegral}
\end{align}
In the second line Feynman parameters $x_1,\ldots,x_N$ have been introduced and the momentum integration has been carried out. The symbols $s_1,\ldots,s_m$ represent invariants or masses and $N_v = \sum_{j=1}^N v_j$. The functions $\mathcal{U}$ and $\mathcal{F}$ are the 1st and 2nd Symanzik polynomials respectively.

After eliminating  the $\delta$-distribution, we obtain integrals which may contain (overlapping) singularities which appear when some of the Feynman parameters vanish simultaneously. Depending on their origin, the singularities which are not integrable correspond to either UV or IR divergences of the Feynman integral. Sector decomposition maps each integral onto a sum of integrals, belonging to so-called sectors $k$,
\begin{equation}
G = (-1)^{N_\nu} \frac{\Gamma(N_\nu-LD/2)}{\prod_{j=1}^N \Gamma(\nu_j)} \sum_k G_k,
\end{equation}
where
\begin{equation}
G_{k} = \int_0^1 \left(\prod_{j=1}^{N-1} \mathrm{d}x_j x_j^{a_{jk}-b_{jk}\epsilon} \right) \frac{\mathcal{U}_{k}(x_1,\ldots,x_{N-1})^{N_\nu - (L+1)D/2}}{\mathcal{F}_{k}(x_1,\ldots,x_{N-1}; s_1,\ldots,s_m)^{N_\nu-LD/2}}. \label{eq:sector}
\end{equation}
Here the functions $\mathcal{U}_{k}$ and $\mathcal{F}_{k}$ are polynomials with a non-zero constant term. In the Euclidean region, defined by all invariants being negative and all masses being positive, they are finite and non-zero everywhere in the integration domain. The singular behaviour of the integrand of each sector can therefore be read off from the exponents $a_{jk}$ and $b_{jk}$. In this form, the integrals can be expanded in $\epsilon$ such that at each order the integrand contains no singularities in the Euclidean region. The integrand is bounded in the integration domain and can be numerically integrated\footnote{For physical kinematics we also use contour deformation~\cite{Soper:1999xk,Binoth:2005ff,Nagy:2006xy,Borowka:2012yc}. This can introduce pinch or end-point singularities near to thresholds. To obtain the differential cross-section we evaluate integrals very close to but not exactly at threshold. For this project no special treatment of thresholds was necessary but numerical convergence was considerably harder to achieve near to threshold, a fact that is reflected in the run time required to obtain a given precision.}.

In the Euclidean region, the singularities of Feynman integrals can alternatively be resolved by choosing a (quasi-)finite basis of integrals~\cite{Panzer:2015ida,vonManteuffel:2014qoa}.  Relations between Feynman integrals (for example integration-by-parts identities~\cite{Tkachov:1981wb,Chetyrkin:1981qh} and dimension shift identities~\cite{Tarasov:1996br}) are used to express integrals in terms of a sum of (quasi-)finite integrals, free of sub-divergences, multiplied by coefficients which contain poles in $\epsilon$ corresponding to the sub-divergences of the original integral. A procedure to do this has already been implemented in \texttt{REDUZE}. 

It has been observed that using a (quasi-)finite basis of integrals along with sector decomposition, as done for $gg \rightarrow HH$, can improve the precision obtained by Monte Carlo (MC) integrators for a given number of sampling points~\cite{vonManteuffel:2015gxa}. In principle, it is not necessary to perform sector decomposition when using a (quasi-)finite basis of integrals, however, integrators such as VEGAS~\cite{Lepage:1977sw,Hahn:2004fe,Hahn:2014fua} and the Rank 1 Shifted Lattice (R1SL), introduced in \sect{sec:r1sl}, typically do not perform well without it. One reason for this is that finite integrals prior to sector decomposition can contain integrable singularities when some Feynman parameters vanish which can spoil the performance of numerical integrators.

Our framework contains an automated interface to \texttt{SecDec}~\cite{Carter:2010hi,Borowka:2012yc,Borowka:2015mxa}, a program that implements sector decomposition. 
Typically \texttt{SecDec} has been used to compute individual integrals to a pre-specified precision. However, here we are interested not in individual integrals but in the amplitude, which consists of a sum of integrals multiplied by coefficients. Thus, rather than obtaining individual integrals with a given precision we wish to obtain the amplitude to a given precision. This was achieved by producing a library containing all sectors and epsilon orders of all integrals which is called by a single amplitude level program. This program dynamically adjusts the precision goal of any particular sector or epsilon order only at runtime until a pre-specified precision for the amplitude is reached, see Ref.~\cite{mkproceedings} for details.

\section{Rank-1 Shifted Lattices \& Implementation} \label{sec:r1sl}

\begin{figure}
\begin{center}
\includegraphics[width=0.65\textwidth]{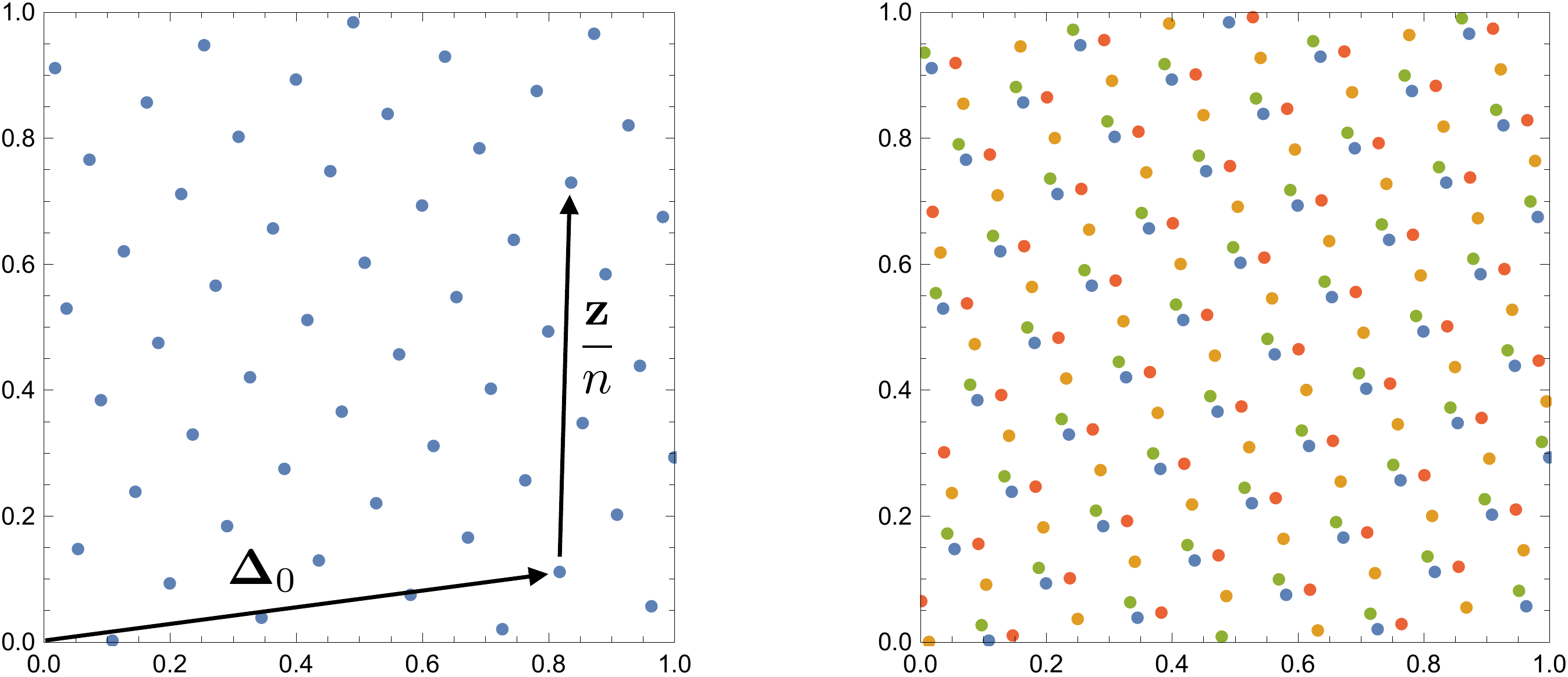}
\end{center}
\caption{(Left panel) An $s=2$ dimensional R1SL with $n=55$ points, generating vector $\mathbf{z}=(1,34)$ and random shift $\boldsymbol{\Delta}_0$. (Right panel) A R1SL produced with three additional random shifts, as used to estimate the mean-square error.}
\label{fig:lattice}
\end{figure}

We wish to numerically compute a multiple integral of a function $f : \mathbb{R}^s \rightarrow \mathbb{R}$ or $f : \mathbb{R}^s \rightarrow \mathbb{C}$ over an $s$-dimensional unit hypercube $[0,1]^s$,
\begin{equation}
I_s [ f ] \equiv \int_{[0,1]^s} \mathrm{d} \mathbf{x} \ f(\mathbf{x}) \equiv \int_0^1 \mathrm{d} x_1 \cdots \mathrm{d} x_s \ f(x_1,\ldots,x_s).
\end{equation}
R1SL are simple QMC cubature rules that have been studied and applied with great success to several numerical integration problems, see Ref.~\cite{ANU:8877392} for a review. In many cases they can be shown to give worst-case root-mean-square errors that scale with the number of (lattice) points arbitrarily close to $\mathcal{O}(n^{-1})$ as compared to MC integrators which scale like $\mathcal{O}(n^{-1/2})$. They were used to numerically integrate sector decomposed Feynman integrals in Ref.~\cite{Li:2015foa} and found to perform extremely well. We have implemented a R1SL integrator and used it to compute all sector decomposed integrals appearing in our problem. 

An unbiased estimate $\bar{Q}_{s,n,m}[f]$ of the integral $I_s[f]$ can be obtained from the R1SL cubature rule~\cite{ANU:8877392},
\begin{align}
&I_s[f] \approx \bar{Q}_{s,n,m}[f] \equiv  \frac{1}{m} \sum_{k=0}^{m-1} Q_{s,n}^{(k)}[f], &
&Q_{s,n}^{(k)}[f] \equiv \frac{1}{n} \sum_{i=0}^{n-1} f \left( \left\{ \frac{i \mathbf{z}}{n} + \boldsymbol{\Delta}_k \right\} \right).& \label{eq:lattice}
\end{align}
The estimate depends on $n$ the number of lattice points and $m$ the number of random shifts. The $\boldsymbol{\Delta}_k \in [0,1)^s$ are $s$-dimensional vectors with components consisting of independent uniformly distributed random real numbers in the interval $[0,1)$. The generating vector $\mathbf{z} \in \mathbb{Z}^s$ is an $s$-dimensional vector of integers which have no factor in common with $n$. The curly braces indicate that the fractional part of each component is taken, such that all arguments of $f$ remain in the interval $[0,1)$.  The lattice of points generated by such a rule is depicted in \fig{fig:lattice}. An unbiased estimate of the mean-square error can be obtained from the random shifts of the lattice according to
\begin{equation}
\mathrm{MSE} [ \bar{Q}_{s,n,m} [f] ] \approx \frac{1}{m(m-1)} \sum_{k=0}^{m-1} ( Q_{s,n}^{(k)}[f] - \bar{Q}_{s,n,m}[f] )^2.
\end{equation}

Lattice rules perform particularly well for continuous and smooth functions which are 1-periodic with respect to each variable. Sector decomposed functions are typically continuous and smooth but not periodic. However, they can be periodized by a suitable change of variables $\mathbf{x}=\phi(\mathbf{y}) = (\phi(y_1),\ldots,\phi(y_s))$. In practice we apply a Korobov transform
\begin{align}
&\phi(y) = \int_0^y \mathrm{d} u \ \omega(u),& 
&\omega(u)= \frac{u^\alpha(1-u)^\alpha}{\int_0^1 \mathrm{d} u \  u^\alpha (1-u)^\alpha},& \label{eq:korobov}
\end{align}
with $\alpha = 3$. Several other Korobov transforms $(\alpha \neq 3)$ and Sidi transforms were also considered but found to give worse performance overall. 

The performance of the R1SL depends strongly on the choice of generating vector $\mathbf{z}$. 
We use the so-called Component-By-Component (CBC) construction~\cite{Nuyens:2006a,nuyens2006fast} as implemented in \texttt{Matlab} in Ref.~\cite{Nuyens2006}. This procedure creates generating vectors that are extensible in $s$ but not $n$. In our implementation we have precomputed a table of generating vectors for $s=6$, the maximum dimension appearing in our problem, which generate lattices consisting of a fixed prime number of points. Concretely, we use lattice rules for Korobov spaces, in the notation of Ref.~\cite{Nuyens2006} they are produced with varying kernel $\Omega(x) = 2 \pi^2 ( x^2-x+1/6)$, product weights $\gamma_i = 1/s$, and parameters $\beta_i = 1$ for $i = 1,\ldots,s$.

\begin{figure}
\begin{center}
\includegraphics[width=.35\textwidth,angle=-90]{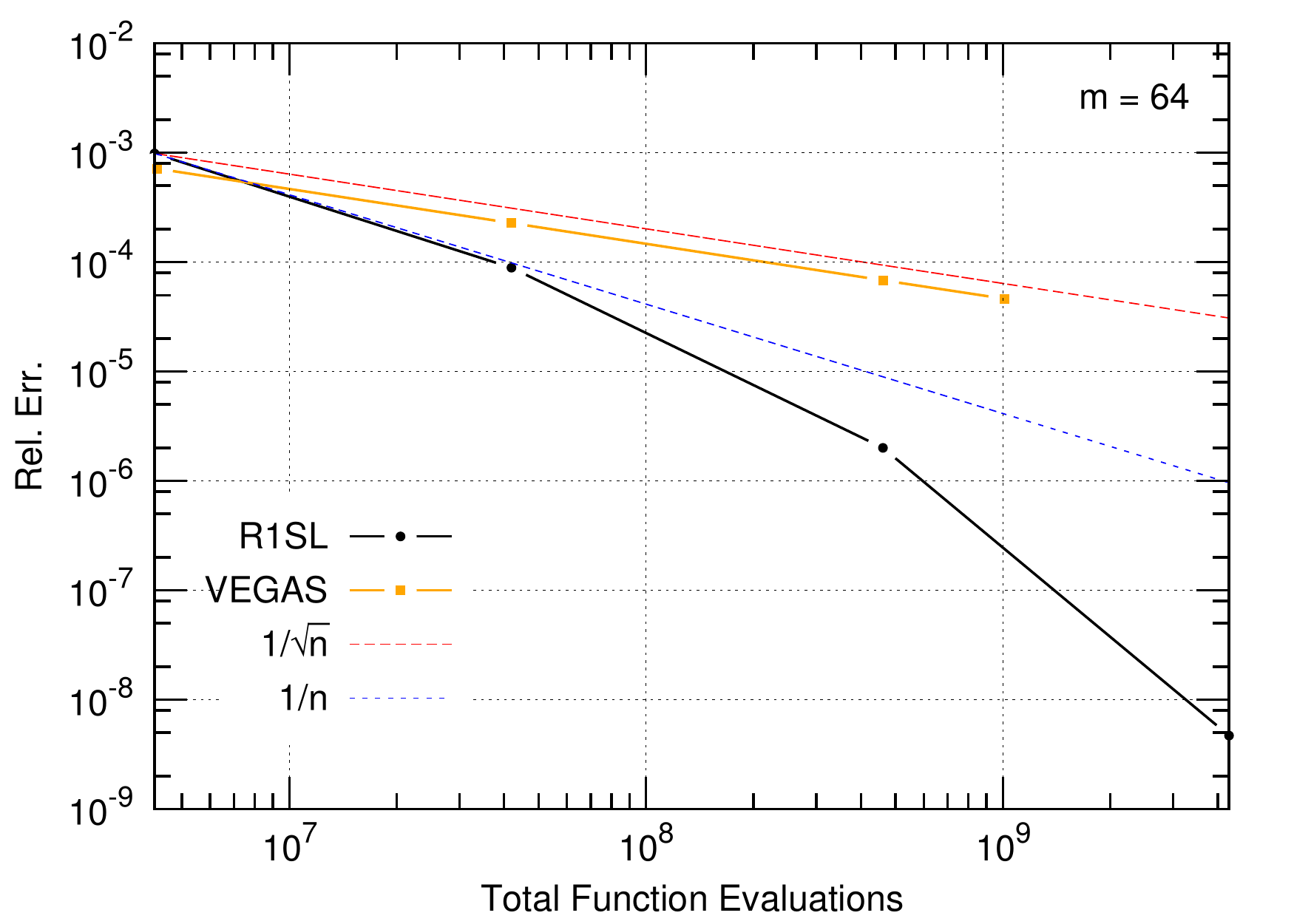}
\includegraphics[width=.35\textwidth,angle=-90]{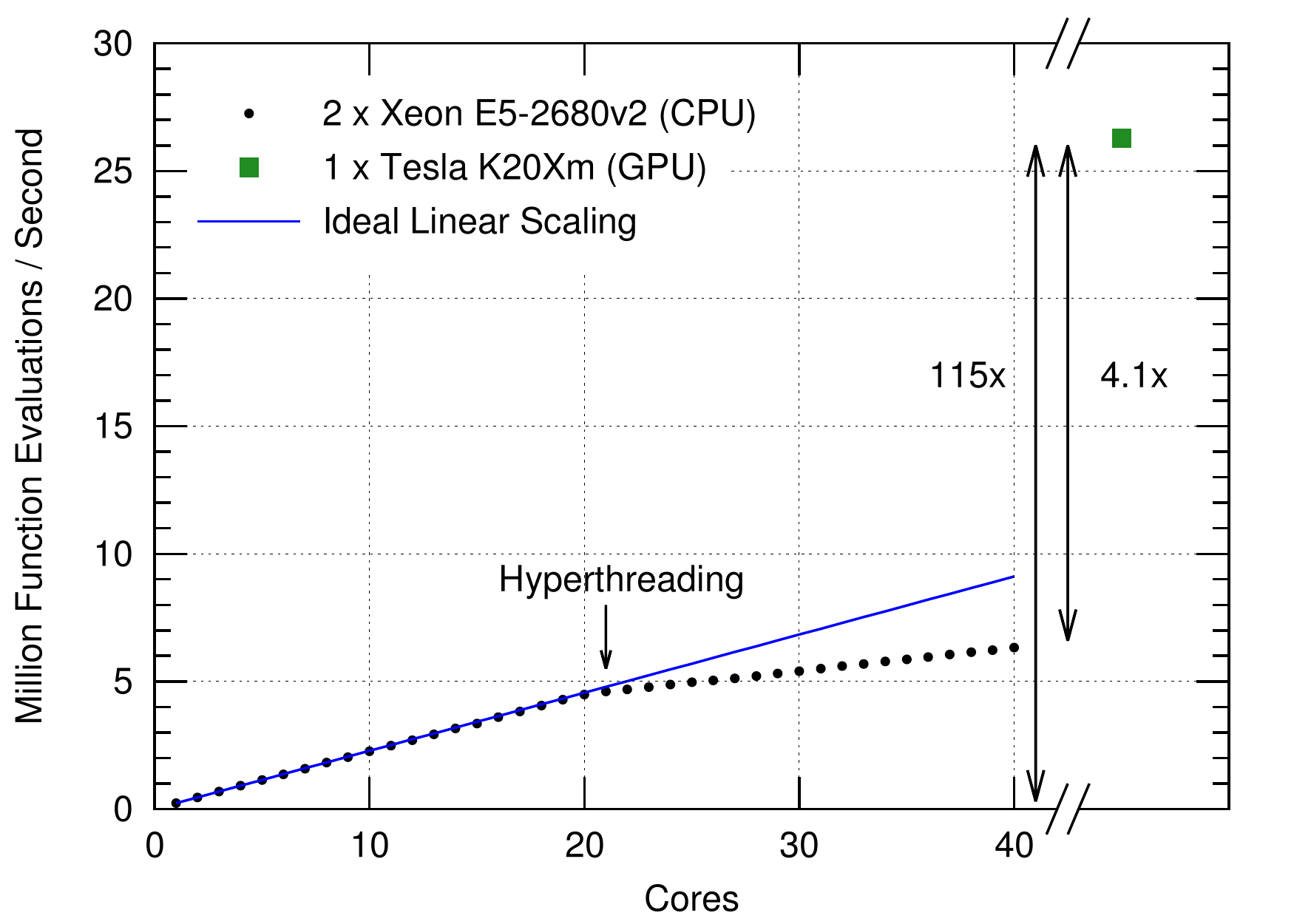}
\end{center}
\caption{(Left panel) The relative error achieved by our R1SL integrator compared to VEGAS for one sector of a 7 propagator 2-loop 4-point integral versus the number of function evaluations used. To fairly compare VEGAS to the R1SL we consider the total number of function evaluations used, for the R1SL this is $n \times m$. Also shown is the expected $\mathcal{O}(n^{-1/2})$ MC scaling and the worst-case $\mathcal{O}(n^{-1})$ R1SL scaling. (Right panel) The number of function evaluations per second achieved by our parallel R1SL implementation versus the number of CPU cores used and the the performance on a GPU.}
\label{fig:implementation}
\end{figure}

In a typical application the number of lattice points $n$ is taken to be large, we use between $10^4$--$10^9$ lattice points. The number of random shifts $m$ is much smaller, typically between 10--50, we use 20 shifts. Our implementation computes all integrals with the smallest lattice $n=65521$. If sufficient precision has not been reached, the integral is then recomputed with an entirely new, larger, lattice. A comparison of the relative error achieved using our R1SL integrator and VEGAS for one sector of a 7 propagator 2-loop 4-point integral appearing in our problem is shown in \fig{fig:implementation}. VEGAS is seen to exhibit MC scaling whilst the R1SL exhibits better than $\mathcal{O}(n^{-1})$ scaling. For moderate numbers of points, $10^7$, VEGAS performs better than the R1SL due to variance reduction, but, for larger $n$ the R1SL performs better due to its scaling. Also worth noting is that whilst VEGAS delivers almost perfect MC scaling the R1SL often performs better than its worst-case error, this behaviour was observed for many of the functions appearing in our project.

The precision reached by the R1SL in a given time is limited by the number of function evaluations.  
In \fig{fig:implementation} we show the number of function evaluations per second achieved by our implementation which uses \texttt{OpenCL 1.1} to numerically integrate using CPUs and/or GPUs. The scaling with the number of CPU cores is linear with (almost) unit gradient up to the number of physical cores then is reduced beyond that, as expected, where hyper-threading is used. To the far right the number of function evaluations per second  achieved by a single \textsc{Nvidia Tesla K20Xm} GPU is shown. The vertical arrows indicate the speed-up obtained by using one GPU compared to one or all cores of two \textsc{Xeon E5-2680v2} CPUs.

\section{Phase-Space Sampling \& Timings} \label{sec:ps}


Due to the use of numerical integration, evaluation of the virtual amplitude for a given phase-space point requires many floating point operations and is consequently slow. It is therefore critically important to decrease as far as possible the number of phase-space points that must be evaluated to obtain quantities of interest. The phase-space parametrisation was implemented by hand and is limited to $2 \rightarrow 2$ or $2 \rightarrow 3$ processes with up to 2 massive particles in the final state.

Unweighted events for the virtual amplitude were generated according to the following two-step procedure: the VEGAS algorithm was applied to the LO matrix element and $\mathcal{O}(10^5)$ events were computed, using these events unweighted events were then generated using the accept/reject method. In total $\mathcal{O}(3 \cdot 10^4)$ events were selected. After this procedure any event exactly reproduces the LO total cross section. The events do not provide a perfect sampling of the phase-space for NLO but, given that the shape of the differential cross-section is not too different from LO, one expects that these events yield a reasonable sampling.

A total of 666 events were computed at NLO. With these events it was possible to obtain the total cross-section with a statistical uncertainty due to the number of events of 0.3$\%$ with an additional uncertainty of 0.1$\%$ coming from the numerical computation of the Feynman integrals. In particular it is worth nothing that events close to the phase space boundary and very close to threshold are computed. One event near to threshold was excluded due to poor convergence of the numerical integrals. Note that no grids are used either for the Feynman integrals or for the events.

Our numeric computations were performed using approximately 16 Dual \textsc{Nvidia Tesla K20X} GPU nodes. Each node computed a different phase-space point and the results were then used later to obtain the total cross-section, invariant mass and $p_T$ distributions on a regular desktop or laptop. The median GPU time per phase-space point was 2 hours with a total of 4680 GPU hours used to compute the events. The total wall-clock time was 6 days. Different cuts, PDFs and even hadron centre-of-mass energies can be selected without recomputing events. Further phase-space points can also be computed at a later date, if necessary, to obtain higher precision for some observable.  

\section{Conclusion}

We have presented some details of our framework which has been used to generate and compute the double Higgs boson production amplitude at 2-loop. It utilises publicly available integral reduction programs and, in lieu of a general master integral library, is interfaced to \texttt{SecDec} for automatic numerical evaluation of entire amplitudes.

Some of the key advances which enabled this calculation were the use of the (quasi-)finite basis, QMC integration and techniques to reduce the number of phase-space points needed to compute observables of interest. Also important was efficiently using compute resources, including a GPU cluster, to obtain the amplitude with the necessary precision. It is anticipated that these techniques will be applicable to other processes, particularly $2 \rightarrow 2$ processes with internal and external masses. 

\section*{Acknowledgements}

I would like to thank Sophia Borowka, Nicolas Greiner, Gudrun Heinrich, Matthias Kerner, Johannes Schlenk, Ulrich Schubert, and Tom Zirke for the fruitful collaboration during this work. 
SPJ is supported by the Research Executive Agency (REA) of the European Union under the Grant Agreement PITN-GA2012316704 (HiggsTools).
We gratefully acknowledge support and resources provided by the Max Planck Computing and Data Facility (MPCDF).

\end{document}